\documentclass[twocolumn,showpacs,preprintnumbers,amsmath,amssymb]{revtex4}
\usepackage{graphicx}
\usepackage{dcolumn}
\usepackage{putexPRL}

\begin{document}

\preprint{PUPT-2275}

\title{The gravity dual to a quantum critical point with spontaneous symmetry breaking}

\author{Steven S. Gubser}\email{ssgubser@Princeton.EDU}
\author{F\'abio D. Rocha}\email{frocha@Princeton.EDU}
\affiliation{Joseph Henry Laboratories, Princeton University, Princeton, NJ 08544}

\date{July 2008}

\begin{abstract}
We consider zero temperature solutions to the Abelian Higgs model coupled to gravity with a negative cosmological constant.  With appropriate choices of parameters, the geometry contains two copies of anti-de Sitter space, one describing conformal invariance in the ultraviolet, and one in the infrared.  The effective speed of signal propagation is smaller in the infrared.  Green's functions and associated transport coefficients can have unusual power law scaling in the infrared.  We provide an example in which the real part of the conductivity scales approximately as $\omega^{3.5}$ for small $\omega$.
\end{abstract}


\maketitle

In \cite{Herzog:2007ij,Hartnoll:2007ih,Hartnoll:2007ip}, it was proposed that $AdS_4$ black holes could be compared to the pseudogap state of high~$T_c$ materials, along the lines of the gauge-string duality \cite{Maldacena:1997re,Gubser:1998bc,Witten:1998qj}.  This proposal hinges on the hypothesis that the properties of the pseudogap are largely controlled by a quantum critical point.  Whether or not this hypothesis is correct, it is clearly desirable to understand better the general proposal that an $AdS_4$ vacuum could describe a quantum critical point.  For a system to exhibit quantum critical behavior, it must have massless excitations.  That usually means a linear dispersion relation, with some characteristic speed $v$ less than the speed of light, and an associated Lorentz group.  Quantum criticality is often (though not necessarily) characterized by relativistic conformal symmetry, which is an enlargement of the Lorentz group: for example, from $SO(2,1)$ to $SO(3,2)$ for theories in two plus one dimensions.  And $SO(3,2)$ is the isometry group of $AdS_4$.  Symmetries of a quantum critical point are emergent in the sense that they characterize the infrared physics of a medium.

In a theory with a gravity dual, how could an $SO(2,1)$ Lorentz symmetry, or a $SO(3,2)$ conformal symmetry, arise as an emergent symmetry at low energies?  Consider the line element
 \eqn{MetricAnsatz}{
  ds^2 = e^{2A} (-h dt^2 + dx^2 + dy^2) + {dr^2 \over h} \,,
 }
where $A$ and $h$ are functions of $r$.  Although the speed of light in the gravity dual is always determined by the structure of null vectors in the local tangent space, the effective speed of transmission of signals in the field theory (say in the $x$ direction) is the coordinate speed $dx/dt$ corresponding to a null vector $v^\mu = (1,\sqrt{h(r)},0,0)$ \footnote{Our convention will be to denote bulk Lorentz indices as $\mu$ or $\nu$, whereas indices in the ${\bf R}^{2,1}$ directions will be noted $m$ or $n$.}.  That is, 
 \eqn{veffDef}{
  v_{\rm eff}(r) = \sqrt{h(r)} \,.
 }
Although the value of $v_{\rm eff}(r)$ can be changed by scaling $t$ and/or $x$ (corresponding to choosing different overall units for time and space), ratios of $v_{\rm eff}(r)$ at different values of $r$ are diffeomorphism-invariant.

In order to have $SO(2,1)$ Lorentz symmetry emerge in the infrared, one needs $h$ to approach a constant in the region where $A \to -\infty$.  (This region is generally understood to correspond to infrared physics in a gravity dual.)  In order for $SO(3,2)$ conformal symmetry to emerge, one must also have $A$ approach a linear function of $r$ in the infrared region, so that the geometry is asymptotically $AdS_4$.  Similar remarks apply in the ultraviolet region, where $A \to +\infty$: Lorentz symmetry arises if $h \to \hbox{constant}$, and conformal symmetry arises if $h \to \hbox{constant}$ and $A$ is asymptotically linear.  Thus we may envision domain wall geometries, where an ultraviolet and infrared geometry, each possessing Lorentzian or conformal symmetry, are separated by a finite region possessing neither.  As we will explain, Einstein's equations together with the null energy condition imply that $v_{\rm eff}(r)$ is a monotonic function of $r$, such that $v_{\rm eff}$ is never greater in the infrared than it is in the ultraviolet.

The domain walls we envision are reminiscent of geometries dual to renormalization group (RG) flows from a UV fixed point to an IR fixed point, but there is one crucial difference: whereas holographic RG flows describe a vacuum configuration of a dual field theory, and therefore respect a Lorentz group that does not change from the UV to the IR, the geometries we aim to construct are supposed to preserve a different Lorentz group in the IR than in the UV---different because $v_{\rm eff}$ is different.

For definiteness, we will focus on an example where there is not only Lorentz invariance, but also conformal invariance both in the ultraviolet and in the infrared.  The action we are going to study is the one proposed in \cite{Gubser:2008px} for describing black holes that superconduct at finite temperature:
 \eqn{AbelianHiggs}{
  S &=  
   \int d^4 x \, \sqrt{g} \Bigg[ 
    R - {1 \over 4} F_{\mu\nu}^2 
     - |\partial_\mu \psi - i q A_\mu \psi|^2 - 
     V(\psi) \Bigg]
 }
where $V(\psi)$ depends only on $|\psi|$.  Crucially, we assume that $V(\psi)$ has a negative local maximum at $\psi=0$ and a gauge-equivalent family of minima at some value $|\psi| = \psi_{\rm IR}$.  The simplest example of a suitable potential is
 \eqn{ChooseV}{
  V(\psi) = -{6 \over L^2} + m^2 |\psi|^2 + {u \over 2} |\psi|^4 \,,
 }
where $m^2 < 0$ and $u>0$.  Then
 \eqn{psiIR}{
  \psi_{\rm IR} = \sqrt{-m^2 \over u} \,.
 }
We can choose coordinates such that the infrared limit of the metric is, call it $AdS_{\rm IR}$, is
 \eqn{dsIR}{
  ds_{\rm IR}^2 = e^{2r/L_{\rm IR}} (-dt^2 + dx^2 + dy^2) + dr^2 \,,
 }
where $L_{\rm IR}$ is defined through the equation
 \eqn{LIRdef}{
  -{6 \over L_{\rm IR}^2} = V(\psi_{\rm IR}) \,.
 }
We need to deform $AdS_{\rm IR}$ at large $r$ in order to get it to match onto the ultraviolet geometry, call it $AdS_{\rm UV}$.  The scalar provides such a deformation: if we assume that $\psi$ is real everywhere, then
 \eqn{psiSet}{
  \psi(r) = \psi_{\rm IR} + a_\psi e^{(\Delta_{\rm IR}-3)r/L_{\rm IR}}
 }
solves the linearized equation of motion for $\psi$ around the background \eno{dsIR}, where $\Delta_{\rm IR}$ is the larger root of
 \eqn{DeltaIR}{
  \Delta_{\rm IR} (\Delta_{\rm IR} - 3) = 
    {1 \over 2} V''(\psi_{\rm IR}) L_{\rm IR}^2
 }
and $a_\psi$ is an undetermined coefficient.

If we solved the equations of motion following from \eno{AbelianHiggs} with the gauge field set to zero and with infrared asymptotics as described in \eno{dsIR} and \eno{psiSet}, we would obtain a holographic RG flow.  A single Lorentz symmetry would be preserved throughout because $\psi$ is a scalar.  To affect the Lorentz symmetry that acts on $(t,x,y)$ but not the rotation symmetry that acts on $(x,y)$, we have to turn on the $t$ component of the gauge field: $A_\mu dx^\mu = \Phi dt$ for some $\Phi(r)$, which must vanish as $r \to -\infty$ because there is a degenerate Killing horizon there where the norm of $dt$ diverges.  (The dual statement is that a quantum critical point whose characteristic velocity is less than the speed of light requires the presence of matter.)  The infrared behavior of $\Phi$ should satisfy the linearized equations of motion for $\Phi$ in the background \eno{dsIR}.  The solution that vanishes as $r\to -\infty$ is
 \eqn{PhiIR}{
  \Phi(r) = \phi_0 e^{(\Delta_{\rm \Phi}-1)r/L_{\rm IR}} \,,
 }
where $\Delta_\Phi$ is the larger root of
 \eqn{DeltaPhi}{
  \Delta_\Phi (\Delta_\Phi-1) = 2 q^2 \psi_{\rm IR}^2 L_{\rm IR}^2
 }
and $\phi_0$ is an undetermined coefficient.  We assume $\phi_0 > 0$, but there is a ${\bf Z}_2$ symmetry between configurations with positive and negative $\phi_0$.  The right hand side of \eno{DeltaPhi} is proportional to the square of the mass of the photon in the symmetry-breaking $AdS_{\rm IR}$ geometry.

The equations of motion following from the action~\eno{AbelianHiggs} are
 \begin{eqnarray}
  A'' &=& -{1 \over 2} \psi'^2 - 
    {q^2 \over 2h^2 e^{2A}} \Phi^2 \psi^2  \label{Aeom} \\
  h'' + 3 A' h' &=& e^{-2A} \Phi'^2 + 
    {2q^2 \over h e^{2A}} \Phi^2 \psi^2  \label{heom} \\
  \Phi'' + A' \Phi' &=& {2q^2 \over h} \Phi \psi^2  \label{Phieom} \\
  \psi'' + \left( 3A' + {h' \over h} \right) \psi' &=& 
    {1 \over 2h} V''(\psi) - 
      {q^2 \over h^2 e^{2A}} \Phi^2 \psi \,,\quad  \label{psieom}
 \end{eqnarray}
and there is a first order constraint, which if satisfied at one value of $r$ must hold everywhere, provided the equations of motion (\ref{Aeom}-\ref{psieom}) are also satisfied:
 \eqn{constraint}{
  &h^2 \psi'^2 + e^{-2A} q^2 \Phi^2 \psi^2 - 
   {1 \over 2} h e^{-2A} \Phi'^2 - 2 hh' A'  \cr
   &\quad\qquad{} - 6 h^2 A'^2 -
   h V(\psi) = 0 \,.
 }
The left hand side of \eqref{Aeom} is proportional to $G^t_t - G^r_r$, where $G_{\mu\nu}$ is the Einstein tensor.  So the right hand side of \eqref{Aeom} is proportional to $T^t_t - T^r_r$.  This quantity is evidently negative for the theory \eno{AbelianHiggs}.  If one considers a more general matter theory coupled to gravity, $T^t_t-T^r_r$ must still be non-positive provided the null energy condition is obeyed: $T_{\mu\nu} \xi^\mu \xi^\nu \geq 0$ for null $\xi^\nu$.  (The symmetries of the metric ansatz demand that $T_{\mu\nu}$ is diagonal.)  This shows that the argument of \cite{Freedman:1999gp} demonstrating the holographic $c$-theorem extends to this case.  Similarly, the left hand side of \eqref{heom} is proportional to $G^x_x - G^t_t$, so the right hand side---call it $s(r)$---must be non-negative of the null energy condition is obeyed.  We can formally solve \eno{heom} in terms of $s(r)$:
 \eqn{FoundH}{
  h(r) = 1 + \int_{-\infty}^r dr_1 \, e^{-3A(r_1)}
    \int_{-\infty}^{r_1} dr_2 \, e^{3A(r_2)} s(r_2) \,.
 }
There are no free integration constants in \eno{FoundH} because we assume $h \to 1$ in the infrared, which implies $e^{3A(r)} h'(r) \to 0$ there as well.  We learn from \eno{FoundH} that $h(r)$, and hence $v_{\rm eff}(r)$, are monotonically increasing functions of $r$, as we claimed earlier.  Note that we do not need to assume that there is conformal invariance in the UV or the IR to obtain \eno{FoundH}.

Before exhibiting an explicit, numerical solution to \eno{Aeom}-\eno{constraint}, let's count the scaling symmetries and parameters that characterize a solution.  The upshot of the discussion will be that, given a definite potential, an extremal solution of the form \eno{MetricAnsatz}, with suitable asymptotic behaviors prescribed for the various fields involved, is essentially unique.

Equations (\ref{Aeom}-\ref{constraint}) have two scaling symmetries that are summarized in Table~\ref{tablesymm}, in which assigning a charge $\alpha$ to a quantity $X$ means that $X \rightarrow \lambda^\alpha X$.
 \begin{table}[htb]
  \centering
  \begin{tabular}{c||c|c|c|c|c||c|c|c}
   \hline\hline
    & $dr$ & $e^A$ & $h$ & $\Phi$ & $\psi$ & $\omega$ & $a_x^{(0)}$ & $a_x^{(1)}$ \\
   \hline
    I & 1 & 0 & 2 & 1 & 0 & 1 & 0 & 0 \\
    II & 0 & 1 & 0 & 1 & 0 & 1 & 0 & 1 \\	
   \hline\hline
  \end{tabular}
  \caption{Charges under scaling symmetries of quantities in (\ref{Aeom}-\ref{constraint}) and (\ref{axEom}).}\label{tablesymm}
 \end{table}
Of the eight integration constants in the equations (\ref{Aeom}-\ref{psieom}), one is used up by \eno{constraint}.  Two are used up by insisting $\psi \to \psi_{\rm IR}$ and $\Phi \to 0$ as $r \to -\infty$.  One is used up by insisting that $h$ is finite as $r \to -\infty$ (meaning that the horizon is degenerate, or zero-temperature).  Using scaling symmetries I and II, we can ensure that $h \to 1$ and that $A(r)-r/L_{\rm IR} \to 0$ as $r \to -\infty$, which uses up
two more integration constants and guarantees that the metric in the far infrared takes the form \eno{dsIR}.  The last two integration constants are the parameters $a_\psi$ and $\phi_0$ in \eno{psiSet} and~\eno{PhiIR}.  But by rescaling $x^m \to \lambda x^m$ and shifting $r/L_{\rm IR} \to r/L_{\rm IR} - \log\lambda$ (which preserves that metric ansatz \eno{MetricAnsatz} and the property $A-r/L_{\rm IR} \to 0$ as $r \to -\infty$), we can one can choose any prescribed value for $\phi_0$.  It appears then that there is a one-parameter family of solutions to (\ref{Aeom}-\ref{constraint}), parametrized by $a_\psi$.  However, the asymptotic behavior of $\psi$ near the conformal boundary is constrained once one chooses a lagrangian for the UV field theory.  For the sake of definiteness, we will assume that the appropriate boundary condition on $\psi$ near the conformal boundary is
 \eqn{DirichletPsi}{
  \psi \propto e^{-\Delta_\psi A} \,,
 }
where $\Delta_\psi$ is the larger root of
 \eqn{DeltaPsi}{
  \Delta_\psi (\Delta_\psi-3) = m^2 L^2 \,.
 }
After this constraint is imposed, there can only be discretely many solutions.  There may be none.  If there is more than one solution, probably the only stable one is the one with the fewest nodes of $\psi$.  If there is a stable solution, it represents a quantum critical point.

\begin{figure}
  \centerline{\includegraphics[width=4in]{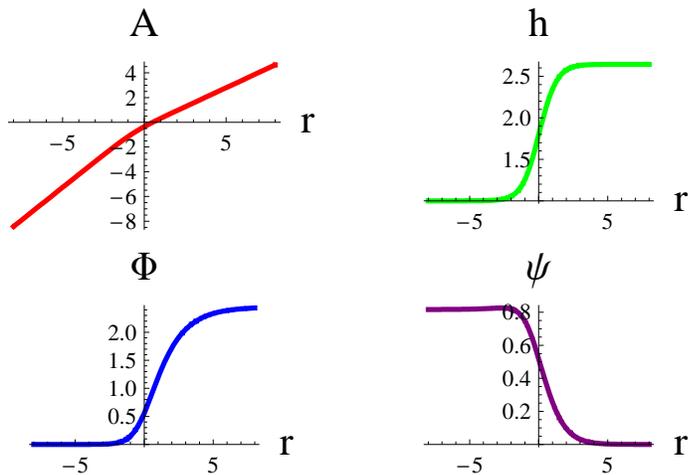}}
  \caption{An example of a solution connecting two AdS vacua with different effective velocities of signal transmission.}\label{DOMAINWALL}
 \end{figure}

To give an explicit example, we used the potential \eno{ChooseV} and made the following choice of parameters:
 \eqn{ChooseParams}{
  L = 1 \qquad q = 2 \qquad m^2 = -2 \qquad u = 3 \qquad \phi_0 = 1 \,.
 }
A simple shooting algorithm suffices to find a solution satisfying both the infrared and ultraviolet asymptotic properties discussed in the previous paragraphs.  We show the result in figure~\ref{DOMAINWALL}.
The solution has $a_\psi = 0.312$, and the ratio of its characteristic velocity in the infrared to the speed of light in the ultraviolet is $v/c = \sqrt{h_{\rm IR}/h_{\rm UV}} = 0.615$.

It is reasonable to expect the transport coefficients of a quantum critical point to have power-law dependence on frequency.  The example that is readiest to hand in our setup is the conductivity.  Consider a complexified perturbation of the gauge field:
 \eqn{aPerturb}{
  A_x = e^{-i\omega t} a_x(r) \,.
 }
Following \cite{Hartnoll:2008vx}, if the leading behavior near the conformal boundary is
 \eqn{axExpand}{
  a_x(r) = a_x^{(0)} + a_x^{(1)} e^{-A(r)} + \ldots \,,
 }
and if $A_x$ is constrained to have purely infalling behavior as one approaches the degenerate Killing horizon of $AdS_{\rm IR}$, then
 \eqn{FoundSigma}{
  \sigma \propto \tilde\sigma \equiv  
    {-i \over \omega} {a_x^{(1)} \over a_x^{(0)}} \sqrt{h_{\rm UV}} \,.
 }
The constant of proportionality in \eno{FoundSigma} is $\omega$-independent.  To relate $\sigma$ to the resistance entering Ohm's Law, one must include factors of the electric charge relating to a weak gauging of the $U(1)$ symmetry of the boundary theory by a boundary gauge field.  We will only be interested in the frequency dependence, so we will not try to track down the constant of proportionality and instead simply compute $\tilde\sigma$. 
The factor of $\sqrt{h_{\rm UV}}$ ensures that this expression is invariant under the scaling symmetries of Table~\ref{tablesymm}.  Because the rotation invariance acting on $(x,y)$ is unbroken, the conductivity is a scalar.  In other words, if we chose to perturb $A_y$ instead of $A_x$, we would get the same answer for the conductivity.

The equation of motion obeyed by $a_x$ is
 \eqn{axEom}{
  & a_x'' + \left( A' + {h' \over h} \right) a_x'  \cr
  &\quad\qquad {} + 
   {1 \over h} \left( {\omega^2 \over h e^{2A}} - 2q^2 \psi^2 - 
     e^{2A} \Phi'^2 \right) a_x = 0 \,.
 }
To derive \eno{axEom}, one must also consider a complexified metric perturbation $\delta g_{tx}$.  This metric perturbation mixes with $a_x$ in the linearized equations, but thanks to a constraint from the $xr$ Einstein equation, $\delta g_{tx}$ can be eliminated altogether from the Maxwell equations, and \eno{axEom} follows more or less immediately.  The charge assignments for $\omega$, $a_x^{(0)}$, and $a_x^{(1)}$ were chosen so that \eno{axExpand} and \eno{axEom} respect the scaling symmetries summarized in Table~\ref{tablesymm}.

Modulo some technical assumptions, it is easy to extract the scaling of $\tilde\sigma$ in the small $\omega$ limit.  First, $\Im\tilde\sigma \sim 1/\omega$, because this is what Kramers-Kronig requires based on the existence of a delta function in $\Re\tilde\sigma$ at $\omega=0$ \footnote{Our first technical assumption is that the continuous part of $\Re\tilde\sigma$ is integrable at $\omega=0$.  This can be checked {\it a posteriori}.}.  In other words, it is a consequence of superconductivity.  Next, $\Re\tilde\sigma$ is related to a conserved flux: if we define
 \eqn{ConservedFlux}{
  {\cal F} = -{he^A \over 2i} a^*_x
    \overleftrightarrow\partial_r a_x
 }
then $\partial_r {\cal F}=0$ follows from \eno{axEom}, and
 \eqn{ReSigma}{
  \Re\tilde\sigma = {L \over \omega \sqrt{h_{\rm UV}}} 
    {{\cal F} \over |a_x^{(0)}|^2}
 }
by direct computation.  Deep in the $AdS_{\rm IR}$ region, where we can ignore deviations from $\psi=\psi_{\rm IR}$ and $\Phi=0$ \footnote{Our second technical assumption is that not only $\Phi \to 0$ as $r \to -\infty$, but also $e^{2A} \Phi' \to 0$, so that the $\Phi'^2$ term in \eno{axEom} can be neglected in comparison with the $\omega^2$ term.  This can be checked given values of the parameters as in \eno{ChooseParams}.}, \eno{axEom} can be solved explicitly:
 \eqn{axNear}{
  a_x = e^{-r/2L_{\rm IR}} H_{\Delta_\Phi-1/2}^{(1)}\left( \omega L_{\rm IR}
    e^{-r/L_{\rm IR}} \right) \,,
 }
where $H_{\Delta_\Phi-1/2}^{(1)}$ is a Hankel function.  Passing \eno{axNear} through \eno{ConservedFlux} gives an $\omega$-independent flux, so to determine the scaling of $\Re\tilde\sigma$ we need only find $a_x^{(0)}$.  The trick that makes this possible is that a solution, call it $a_x(r) = Z_x(r)$, to the $\omega \to 0$ limit of \eno{axEom}, can be combined with \eno{axNear} using the method of matched asymptotic expansions, provided $\omega$ is small.  $Z_x$ should be chosen so that $e^{-(\Delta_\phi-1) r/L_{\rm IR}} Z_x \to 1$ as $r \to -\infty$.  Assume that $\omega$ is small enough so that the radius $r_* \equiv L \log \omega L$ is much less than the radius $r_{\rm IR}$ at which $AdS_{\rm IR}$ is significantly deformed.  In the matching window $r_* \ll r \ll r_{\rm IR}$, one finds by expanding the Hankel function that \footnote{Our third technical assumption is that $\Delta$ is not half an integer.  If it is, then the expansion of $H_{\Delta_\Phi-1/2}$ involves a logarithm, and logarithmic scaling violations may result.} 
 \eqn{axWindow}{
  a_x(r) \approx i \sec\pi\Delta_\phi 
   \left( {\omega L \over 2} \right)^{-\Delta_\Phi+1/2} Z_x(r) \,.
 }
In fact, the approximate equality \eno{axWindow} must hold for all $r \gg r_*$, because the $\omega^2$ term in \eno{axEom} is negligible in this region.  So we conclude
 \eqn{axZeroScaling}{
  a_x^{(0)} = \lim_{r \to \infty} a_x(r) \propto \omega^{-\Delta_\Phi+1/2} \,,
 }
where the constant of proportionality involves $\lim_{r\to\infty} Z_x(r)$, which encodes all the physics at high scales that we can access only through numerical methods.  Plugging \eno{axZeroScaling} into \eno{ReSigma}, one obtains $\Re\tilde\sigma \sim \omega^\delta$ for small $\omega$, with
 \eqn{deltaDef}{
  \delta = 2(\Delta_\Phi-1) \,.
 }
For the parameters indicated in \eno{ChooseParams}, $\Delta_\Phi = {1 \over 2} + \sqrt{101 \over 20}$, resulting in $\delta \approx 3.5$.
We also note that in the high-$\omega$ limit, we expect that only the geometry near the boundary will be relevant for the calculation
of the conductivity. Since this geometry is nearly $AdS_4$ with $\psi=0$, $\tilde \sigma$ should asymptote to its value in that background, which is constant \cite{Herzog:2007ij}: $\tilde \sigma_{AdS_4}=1$ in our conventions.
\begin{figure}[h]
\vspace{7pt}
  \centerline{\includegraphics[width=3.5in]{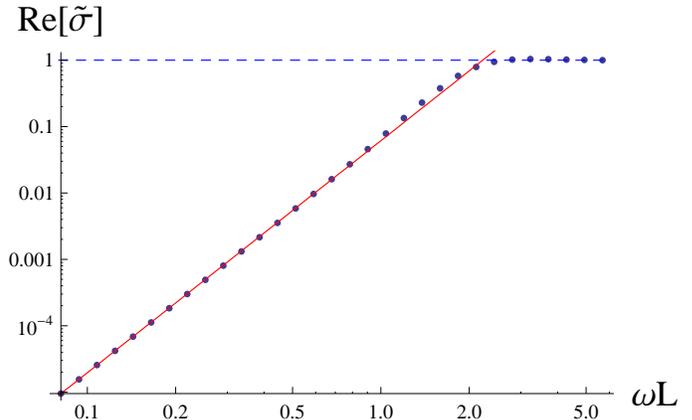}}
  \caption{The real part of $\tilde \sigma$ as function of $\omega L$ for the solution displayed in figure~\ref{DOMAINWALL}.
  The dots show the result of numerical computation while the solid line is the small $\omega$  power-law behavior 
  $\Re\tilde\sigma \sim \omega^\delta$ with the overall constant chosen so the line passes through the first point.
  The dashed line shows the high $\omega$ limit $\tilde \sigma =1$.
  Note that there is an ambiguity in the scale of $\omega$, as the scaling symmetries affect it. In this plot the scale is fixed by 
  using the scaling symmetries to ensure that $h \to 1$ and $A(r) - r/L \to 0$ as $r \to +\infty$.}\label{COLDSIGMA}
 \end{figure}

Backgrounds of the type we have described probably exist in dimensions other than four.  Non-trivial scaling laws in the infrared probably arise for other Green's functions, for example the ones associated with the operator dual to $\psi$.  As we remarked earlier, it would be interesting to generalize to cases where the symmetries displayed by the ultraviolet and/or infrared regions are only the Lorentz group rather than the full relativistic conformal group.  It is also important to determine whether such solutions are zero-temperature limits of regular black hole solutions which are stable and thermodynamically favored over all other finite-temperature configurations.

\begin{acknowledgments}

We thank S.~Pufu for collaboration on finding the perturbation equation \eno{axEom} and A.~Nellore for useful discussions.  The work of S.S.G.~was supported in part by the Department of Energy under Grant No.\ DE-FG02-91ER40671 and by the NSF under award number PHY-0652782.  F.D.R.~was supported in part by the FCT grant SFRH/BD/30374/2006.

\end{acknowledgments}

\bibliographystyle{apsrev}
\bibliography{photon}
\end{document}